\documentclass[RNAAS]{aastex62}
\usepackage{amsmath}

\newcommand{\be}{\begin{equation}}
\newcommand{\ee}{\end{equation}}
\newcommand{\ba}{\begin{eqnarray}}
\newcommand{\ea}{\end{eqnarray}}
\newcommand{\bse}{\begin{subequations}}
\newcommand{\ese}{\end{subequations}}

\begin{document}

\title{On modeling CC85 wind in an Expanding local Box }

\correspondingauthor{Alankar Dutta}
\email{alankardutta@iisc.ac.in, dutta.alankar@gmail.com}

\author[0000-0002-9287-4033]{Alankar Dutta}
\affiliation{Department of Physics \\
Indian Institute of Science \\
Bangalore, Karnataka 560012, India}

\author{Prateek Sharma}
\affiliation{Department of Physics \\
Indian Institute of Science \\
Bangalore, Karnataka 560012, India}
\email{prateek@iisc.ac.in}



\section{Motivation} 
\citet{Chevalier1985} [henceforth CC85] proposed a model for a steady thermalized hot wind expanding radially outward from a central starburst. This can be taken as the simplest model for superbubble feedback due to coalescing supernovae (\citealt{Sharma2014}), which results in a galactic outflow. One of the crucial problems in galactic outflows is whether cold clouds can survive in them. Recent plane-parallel wind tunnel simulations (\citealt{Gronke2018}) suggest that this may be possible if the cloud size is sufficiently large. It is important to answer this question in a more realistic spherical wind. In past, such a spherically expanding wind has been modelled locally as a coordinate expansion (\citealt{Scannapieco2017,Gronke2019}). Mathematically, this is analogous to the use of  comoving coordinates to account for Hubble (cosmological) expansion of the Universe. However, these works do not present the evolution equations explicitly. Moreover, we highlight that, unlike isotropic cosmological expansion, the local coordinate expansion due to a radial wind is anisotropic  occurring only in the angular directions but not radially. The aim of this Note is to present the governing equations explicitly.

\section{Transforming Fluid Equations} 
\begin{figure}
\begin{center}
\includegraphics[scale=0.5,angle=0]{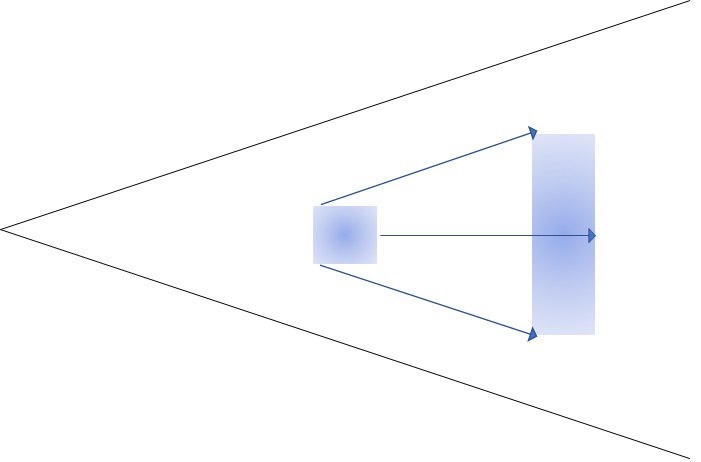}
\caption{A local Cartesian box moving with the CC85 wind only experiences expansion in the directions orthogonal to the wind. \label{fig:1}}
\end{center}
\end{figure}

Consider a small cuboidal box (of size $\ll r$, its radial location) frozen in a steady CC85 wind (see Fig. \ref{fig:1}). We choose local Cartesian coordinates $x, y, z$ along the radial, polar and azimuthal directions respectively. For a wind of constant speed there is no radial expansion of the box but there is expansion in the orthogonal directions attributable to the spherical geometry. This orthogonal expansion can be modelled with a time dependent scale parameter.  

We use the following transformations between the physical (denoted by tilde) and expanding coordinates
\bse \label{eq:coordinates}
\begin{align}
\label{eq:x}
\widetilde{x} &= x, \\
\label{eq:y}
\widetilde{y} &= a(t)y, \\ 
\label{eq:z}
\widetilde{z} &= a (t)z, \\
\label{eq:t}
\widetilde{t} &= t, 
\end{align}
\ese
where $a(t)$ is the time dependent scale factor that accounts for the orthogonal expansion (with respect to the wind direction) in a CC85 wind. The scalar fields are chosen to transform as
\bse \label{eq:field}
\begin{align}
\widetilde{\rho} &= a^{-2} \rho, \\
\widetilde{p} &= a^{-2\gamma}p, \\
\widetilde{\epsilon} &= a^{-2(\gamma-1)}\epsilon.
\end{align}
\ese
The fluid field transformation (Eqs. \ref{eq:field}) is chosen such that the CC85 wind is adiabatic and the equation of state $\rho \epsilon = p/(\gamma - 1)$ remains the same in both coordinates. The velocity transformations are chosen to be
\bse \label{eq:vel_transf}
\begin{align}
\widetilde{v}_x &= v_x, \\
\widetilde{v}_y &= a v_y+\dot{a} y, \\
\widetilde{v}_z &= a v_z+\dot{a} z,
\end{align}
\ese
which accounts for the anisotropic expansion in the orthogonal directions. We note that the choice of transformations (Eqs. \ref{eq:coordinates}-\ref{eq:vel_transf}) is arbitrary but transformations that simplify the evolution equations are preferred. 
Eqs. \ref{eq:coordinates} imply that the partial derivatives transform as
\bse \label{eq:pderv_transf}
\begin{align}
\frac{\partial}{\partial \widetilde{x}} &= \frac{\partial}{\partial x}, \\
\frac{\partial}{\partial \widetilde{y}} &= \frac{1}{a}\frac{\partial}{\partial y}, \\
\frac{\partial}{\partial \widetilde{z}} &= \frac{1}{a}\frac{\partial}{\partial z}, \\
\frac{\partial}{\partial \widetilde{t}} &= \frac{\partial}{\partial t}-\frac{\dot{a}}{a}\left(y\frac{\partial}{\partial y}+z\frac{\partial}{\partial z}\right).
\end{align}
\ese

Using these transformations, and in the absence of gravity, the fluid equations in the locally expanding coordinates become \\
\textbf{Mass conservation equation}:
\be \label{eq:cont}
\frac{\partial \rho}{\partial t}+\frac{\partial}{\partial x} (\rho v_x)+\frac{\partial}{\partial y} (\rho v_y)+\frac{\partial}{\partial z} (\rho v_z) = 0,
\ee
\textbf{Momentum (Euler) equations}:
\bse \label{eq:eul}
\begin{align}
\frac{\partial }{\partial t}(\rho v_x)+\frac{\partial}{\partial x} (\rho v_x^2)+\frac{\partial}{\partial y} (\rho v_x v_y)+\frac{\partial}{\partial z} (\rho v_x v_z) &= -a^{-2(\gamma - 1)}\frac{\partial p}{\partial x}, \label{eq:eul_x}\\
\frac{\partial }{\partial t}(\rho v_y) + \frac{\partial}{\partial x} (\rho v_x v_y)+\frac{\partial}{\partial y} (\rho v_y^2)+\frac{\partial}{\partial z} (\rho v_y v_z)+\left(2 \frac{ \dot{a}}{a}v_y+\frac{\Ddot{a}}{a}y\right)\rho &= -a^{-2\gamma }\frac{\partial p}{\partial y} \label{eq:eul_y},\\
\frac{\partial }{\partial t}(\rho v_z)+ \frac{\partial}{\partial x} (\rho v_x v_z)+\frac{\partial}{\partial y} (\rho v_y v_z)+\frac{\partial}{\partial z} (\rho v_z^2) +\left(2 \frac{\dot{a}}{a}v_z+\frac{\Ddot{a}}{a}z\right)\rho &= -a^{-2\gamma }\frac{\partial p}{\partial z} \label{eq:eul_z},
\end{align}
\ese
\textbf{Entropy equation}:
\be \label{eq:entrp}
\left(\frac{\partial}{\partial t}+v_x \frac{\partial}{\partial x}+ v_y \frac{\partial}{\partial y}+ v_z \frac{\partial}{\partial z}\right)\left(\frac{p}{\rho ^\gamma}\right) = 0.
\ee
The form of the internal energy equation in the frame of the expanding local box is the same as in physical coordinates. For completeness, the total energy equation is given by
\be
\label{eq:tot_energy}
\frac{\partial}{\partial t} \left( \frac{\rho v^2}{2} + \frac{p}{\gamma-1} \right) + \nabla \cdot \left[  \left ( \frac{\rho v^2}{2} + \frac{\gamma p}{\gamma-1} \right) {\bf v}\right] = -(a^{-2(\gamma-1)}-1) v_x \frac{\partial p}{\partial x} - (a^{-2\gamma}-1) \left( v_y \frac{\partial p}{\partial y} + v_z \frac{\partial p}{\partial z}  \right) -\frac{2 \dot{a} \rho}{a}(v_y^2 + v_z^2)  -\frac{\Ddot{a} \rho}{a}(y v_y + z v_z ).
\ee

\section{A constant velocity wind}
The CC85 wind reaches a constant speed at large radii where $\widetilde{\rho} \propto r^{-2}$. This allows us to simplify the expression for the scale factor to
\be
a(t) \equiv \frac{r(t)}{r_0}=1+\frac{v_w t}{r_0},
\ee
where $v_w$ is the constant wind velocity. The scale factor $a(t)$ is chosen to be unity at the initial location of the local box ($r_0$) around the cloud and increases linearly with time. For the cloud crushing problem, a blob is initialized at rest in the lab frame and the cloud-wind interaction is followed in time. For a constant velocity wind, $\dot{a}=v_w / r_0$ and $\Ddot{a}=0$ and the transverse momentum (Eqs. \ref{eq:eul_y} \& \ref{eq:eul_z}) and the total energy (Eq. \ref{eq:tot_energy}) equations simplify to
\bse
\begin{align}
\frac{\partial }{\partial t}(\rho v_y)+\frac{\partial}{\partial x} (\rho v_x v_y)+\frac{\partial}{\partial y} (\rho v_y^2)+\frac{\partial}{\partial z} (\rho v_y v_z)+2\frac{\dot{a}}{a}v_y\rho &= -a^{-2\gamma }\frac{\partial p}{\partial y},\\
\frac{\partial }{\partial t}(\rho v_z)+\frac{\partial}{\partial x} (\rho v_x v_z)+\frac{\partial}{\partial y} (\rho v_y v_z)+\frac{\partial}{\partial z} (\rho v_z^2)+2\frac{\dot{a}}{a}v_z\rho &= -a^{-2\gamma }\frac{\partial p}{\partial z},\\
\frac{\partial}{\partial t} \left( \frac{\rho v^2}{2} + \frac{p}{\gamma-1} \right) + \nabla \cdot \left[  \left ( \frac{\rho v^2}{2} + \frac{\gamma p}{\gamma-1} \right) {\bf v}\right] = -(a^{-2(\gamma-1)}-1) v_x \frac{\partial p}{\partial x} &- (a^{-2\gamma}-1) \left( v_y \frac{\partial p}{\partial y} + v_z \frac{\partial p}{\partial z}  \right) -\frac{2 \dot{a} \rho}{a}(v_y^2 + v_z^2).
\end{align}
\ese
\section{Cloud tracking with frame boost}
In the previous section we presented a convenient choice of variables and accordingly transformed the governing equations. However, solving these equations either in the lab frame or in a frame moving with the wind is not desirable as the cloud, which we want to follow, is initially at rest in the lab frame and eventually moves with  the wind. This is true both in presence and absence of background expansion. \citet{Shin2008,McCourt2015,Gronke2018,Gronke2019} use a cloud tracking scheme to continually switch to the frame in which the cloud is at rest and the cloud-wind interaction region remains within the computational domain for a longer time. This method is only briefly discussed in \citet{Shin2008}. We explain the procedure for implementing this frame boost in some detail here.

At each timestep in the simulation, the computational domain is boosted to follow the cloud. To track the cloud we set a tracer field value of unity in the cloud and zero in the wind. The tracer is passively advected and we track the average cloud velocity in  the $x$- (wind) direction,
\be \label{eq:tracer_vel}
\left<v_x\right> = \frac{\int{\rho C v_x dV}}{\int{\rho C dV}},
\ee
where the tracer (color) is indicated by the variable $C$. If the velocity field everywhere is changed to ${\bf v}({\bf r})-\left<v_x\right> {\bf \hat{x}}$ at every timestep, the frame gets boosted to follow the cloud. This procedure implicitly provides the necessary pseudo-force for transforming to this accelerated frame with a velocity boost of ${\bf \Delta v} = -\left<v_x\right> {\bf \hat{x}} = {\bf a}_{\rm pseudo} \Delta t$ ($\Delta t$ is the computational timestep). Note that there is no need to add an explicit pseudo-force term to the momentum equation.

Another subtlety to note is that the cloud velocity found in this way (Eq. \ref{eq:tracer_vel}) at any timestep is with respect to the boosted frame in the previous timestep. To obtain the velocity in the lab frame, one must add up all the velocity boosts till the current timestep. Similarly, the radial position of the cloud in the lab frame can be obtained by summing up the contributions from each timestep.

\acknowledgments
We would like to thank Max Gronke for useful correspondence. PS acknowledges a Swarnajayanti fellowship (DST/SJF/PSA-03/2016-17) from DST India. PS also thanks the Humboldt Foundation that enabled his sabbatical at MPA where this work was initiated.

\bibliographystyle{aasjournal}
\bibliography{rnaas}

\end{document}